\documentclass[11pt,a4paper,epsf,epsfig,psfrag]{article}
\usepackage{jheppub}
\usepackage{amsmath,graphicx,amsfonts, mathrsfs,amssymb}
\usepackage{amsmath,epsfig}
\usepackage{amssymb,amsfonts}
\usepackage{color}
\usepackage{latexsym}
\usepackage{epsfig}
\usepackage{pdfsync}
\newbox\pippobox

\def\be{\begin{equation}}
\def\ee{\end{equation}}
\def\bea{\begin{eqnarray}}
\def\eea{\end{eqnarray}}

\newcommand{\beq}{\begin{equation}}
\newcommand{\eeq}{\end{equation}}
\newcommand{\beqa}{\begin{eqnarray}}
\newcommand{\eeqa}{\end{eqnarray}}
\newcommand{\beqar}{\begin{eqnarray*}}
\newcommand{\eeqar}{\end{eqnarray*}}

\renewcommand{\eqref}[1]{(\ref{#1})}

\catcode`\@=12

\title{Chiral phase transition of QCD with $N_f=2+1$ flavors from holography}

\author[a]{Danning Li,}
\author[b,c]{Mei Huang}

\affiliation[a]{Department of Physics, Jinan University, Guangzhou 510632, P.R. China}
\affiliation[b]{Institute of High Energy Physics, Chinese Academy of Sciences, Beijing 100049, P.R. China}
\affiliation[c]{Theoretical Physics Center for Science Facilities, Chinese Academy of Sciences, Beijing 100049, P.R. China}

\emailAdd{lidanning@jnu.edu.cn}
\emailAdd{huangm@ihep.ac.cn}

\abstract{Chiral phase transition for three-flavor $N_f=2+1$ QCD with $m_u=m_d\neq m_s$ is investigated in a modified soft-wall holographic QCD model. Solving temperature dependent chiral condensates from equations of motion of the modified soft-wall model, we extract the quark mass dependence of the order of chiral phase transition in the case of $N_f=2+1$, and the result is in agreement with the ``Colombia Plot", which is summarized from lattice simulations and other non-perturbative methods. First order phase transition is observed around the three flavor chiral limit $m_{u/d}=0, m_{s}=0$, while at sufficient large quark masses it turns to be a crossover phase transition. The first order and crossover regions are separated by a second order phase transition line. The second order line is divided into two parts by the $m_{u/d}=m_s$ line, and the $m_s$ dependence of the transition temperature in these two parts are totally contrast, which might indicate that the two parts are governed by different universality classes.}
\keywords{Chiral phase transition, chiral condensate, soft-wall AdS/QCD}

\begin{document}
\maketitle

\section{Introduction}
\label{sec-int}

Quantum Chromodynamics(QCD) is widely accepted as the fundamental theory of strong interactions, and QCD vacuum is characterized by spontaneous chiral symmetry breaking together with color charge confinement. The dynamically generated chiral condensate $\langle \bar{\psi}\psi\rangle$ in the vacuum serves as quarks' dynamical mass and spontaneously breaks the chiral symmetry, which is an exact symmetry of QCD lagrangian when quarks are massless. It is believed that at sufficient high temperature and/or density, quark condensate might be destroyed completely and the spontaneous breaking symmetry would be restored. Understanding the property of chiral phase transition has been an important topic in both non-perturbative QCD and cosmology for decades\cite{Aoki:2006we}.

\begin{figure}[h]
\begin{center}
\epsfxsize=6.5 cm \epsfysize=6.5 cm \epsfbox{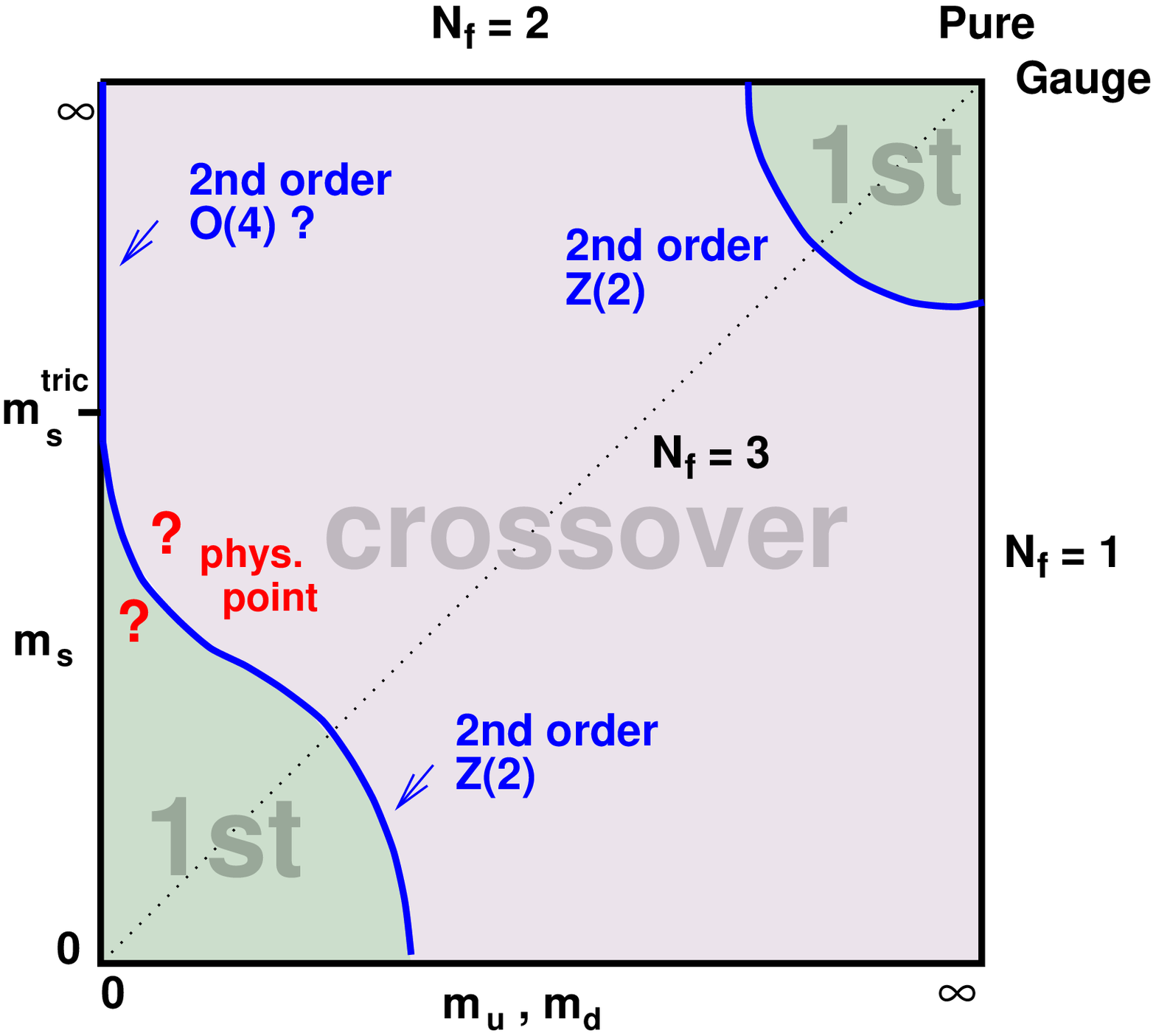}
\hspace*{0.1cm} \epsfxsize=6.5 cm \epsfysize=6.5 cm \epsfbox{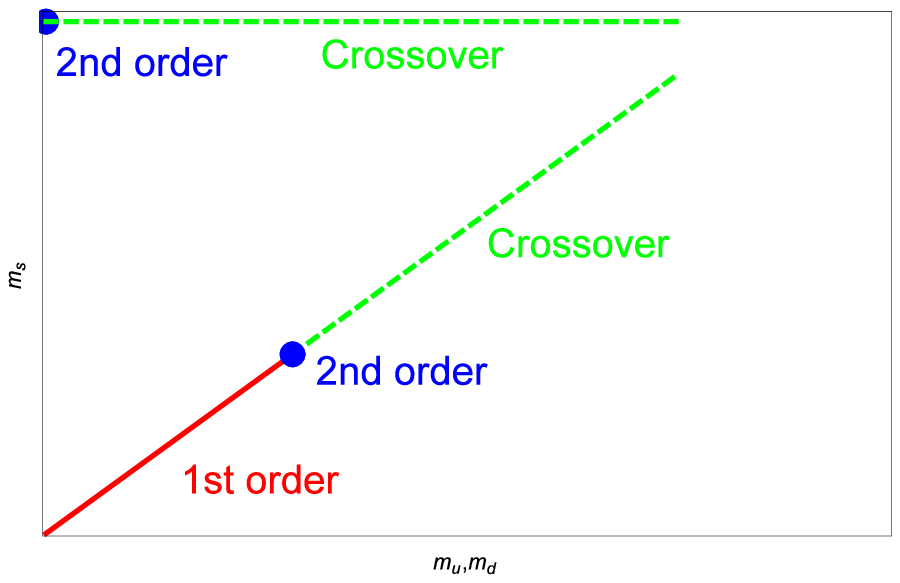}
\vskip -0.05cm \hskip 0.15 cm
\textbf{( a ) } \hskip 6.5 cm \textbf{( b )} \\
\end{center}
\caption{Panel.(a) shows the ``Colombia Plot" which gives expected phase diagram in the quark masses $m_u=m_d, m_s$ plane (Taken from \cite{qcd-phase-diagram}). Panel.(b) gives the prediction of the mass diagram of $N_f=2$ and $N_f=3$ in a modified soft-wall model(Taken from \cite{Chelabi:2015gpc}). }
\label{Colombia-Plot}
\end{figure}

The order of chiral phase transition depends sensitively on the inertial degrees of freedom of the system, such as the number of flavors($N_f$) and the mass of quarks($m_u,m_d$ and $m_s$). Based on theoretical consideration and lattice QCD simulations \cite{qcd-phase-diagram,deForcrand:2006pv,Kanaya:2010qd}, the expected three flavor phase diagram in the quark mass plane, describing quark mass dependence of the order of QCD phase transitions, is summarized in the sketch plot (it is also called ``Columbia Plot") shown in Fig.\ref{Colombia-Plot}(a). In this sketch plot, the whole $m_{u/d}-m_s$ plane are divided into three simply connected region: two first order zones in the bottom left corner and the upper right corner as well as the crossover region in the middle. The upper right corner is near the infinite quark mass limit $m_u=m_d=m_s=\infty$, where the breaking and restoration of $Z_3$ centre symmetry, related to confinement/deconfinement phase transition, are well defined. The bottom left corner is near the chiral limit $m_u=m_d=m_s=0$, where the breaking and restoration of chiral symmetry, related to chiral phase transition, are well defined. In between these two regions, there are no known exact symmetries and the phase transitions are expected to be a continual and rapid transition between different phases, usually named ``crossover". The boundary of these three regions are second order lines, where the phase transitions are expected to be of second order. Furthermore, we noted that the second order line between the bottom left first order region and the crossover region is divided by the $SU(3)$ diagonal line with $m_u=m_d=m_s$ into two parts, the upper one of which is expected to be governed by the $O(4)$ universality class\cite{Pisarski:1983ms} while the lower one of which is expected to be governed by the $Z(2)$ universality class.

Theoretically, the dynamics of QCD near phase transition is non-perturbative, and normal perturbative methods of quantum field theory become invalid here. Lattice QCD has been considered as the most reliable non-perturbative method to study non-perturbative properties of QCD. However, lattice QCD simulations still require further improvements in several aspects, especially the difficulty called sign problem at finite chemical potential, which is waited to be solved in order to extract full understanding on QCD phase diagram. Hence, it is quite necessary to develop other non-perturbative methods. The recent progress of  anti-de Sitter/conformal field theory (AdS/CFT) correspondence and the conjecture of the gravity/gauge duality\cite{Maldacena:1997re,Gubser:1998bc,Witten:1998qj} does provide such a new powerful tool to tackle the strong coupling problem of gauge theory like QCD.

In the framework of holography, by breaking the conformal symmetry of the original AdS/CFT correspondence in different ways, efforts towards realistic holographical description of non-perturbative physics of QCD, such as hadron physics\cite{Erlich:2005qh,Karch:2006pv,TB:05,DaRold2005,D3-D7,D4-D6,SS-1,SS-2,Csaki:2006ji,Dp-Dq,Gherghetta-Kapusta-Kelley,Gherghetta-Kapusta-Kelley-2,YLWu,YLWu-1,Li:2012ay,Li:2013oda,Bartz:2014oba,Colangelo:2008us,Bellantuono:2015fia,Capossoli:2013kb,Capossoli:2015ywa,Capossoli:2016kcr,Capossoli:2016ydo,Chen:2015zhh,Vega:2016gip} and hot/dense QCD matter\cite{Shuryak:2004cy,Tannenbaum:2006ch,Policastro:2001yc,Cai:2009zv,Cai:2008ph,Sin:2004yx,Shuryak:2005ia,Nastase:2005rp,
Janik:2005zt,Nakamura:2006ih,Sin:2006pv,Herzog:2006gh,Gubser-drag,Wu:2014gla,Li:2014dsa,Li:2014hja},  have been made both in top-down approaches and in bottom-up approaches(see Refs.\cite{Aharony:1999ti,Erdmenger:2007cm,deTeramond:2012rt,Kim:2012ey,Adams:2012th} for reviews). Different from top-down approach, bottom-up holographic QCD starts from QCD phenomena and try to build up more realistic holographic models. In this approach, chiral phase transition has been studied in several different models \cite{Iatrakis:2010jb,Jarvinen:2011qe,Alho:2012mh,Colangelo:2011sr,Dudal:2015wfn,Evans:2016jzo,Bartz:2016ufc,Chelabi:2015cwn,Chelabi:2015gpc,Fang:2015ytf,Li:2016gfn}. Most of these studies considered the case with equal quark masses for all quarks. As can be seen from Fig.\ref{Colombia-Plot}(a), it is also interesting to consider the cases when $m_{u/d}\neq m_s$, where the physical point locates. In our previous studies\cite{Chelabi:2015cwn,Chelabi:2015gpc}, based on soft-wall AdS/QCD model \cite{Karch:2006pv}, the mass dependence of chiral phase transition is extracted, as shown in Fig.\ref{Colombia-Plot}(b). The qualitative results for $SU(2)$ and $SU(3)$ cases are in good agreement with the current understanding from Fig.\ref{Colombia-Plot}(a): for two flavor case it starts from a second order phase transition and turns to be crossover at any finite quark mass while for three flavor case it starts from a first order phase transition and only at sufficient large quark masses it turns to be crossover. Furthermore, in \cite{Li:2016gfn}, we extend these studies to finite magnetic field, and find that it can provide good description on inverse magnetic catalysis effect, which was discovered in lattice QCD\cite{Bali:2011qj,Bali:2012zg} and studied in other methods\cite{Fukushima:2012kc, Kojo:2012js, Bruckmann:2013oba, Chao:2013qpa, Fraga:2013ova, Ferreira:2014kpa, Farias:2014eca, Yu:2014sla, Andersen:2014oaa, Ferrer:2014qka, Feng:2014bpa} recently. However, the cases when $m_{u/d}\neq m_s$ have not been examined in this model. Therefore, in this work, we will extend our studies in \cite{Chelabi:2015cwn,Chelabi:2015gpc} to $N_f=2+1$ when $m_l\equiv m_{u}=m_d\neq m_s$ and study the property of chiral phase transition.

The paper is organized as follows. In Sec.\ref{section_model}, we give a short introduction on the model and the numerical method we used, especially on how to introduce quark masses and chiral condensates in the model. Then in Sec.\ref{massdiagram} we show numerical results from our model study, especially the quark mass dependence of the order of the chiral phase transition like Fig.\ref{Colombia-Plot}. Finally in Sec.\ref{sum} a brief discussion will be given.

\section{Soft-wall model in $N_f=2+1$ case}
\label{section_model}

\subsection{Background}
In the original paper of soft-wall model \cite{Karch:2006pv}, the 4D global chiral symmetry $SU(N_f)_L\times SU(N_f)_R$ is promoted to 5D and becomes local gauge symmetry of the following action
\begin{eqnarray}\label{kkssaction}
 S=&&-\int d^5x
 \sqrt{-g}e^{-\Phi}Tr(D_m X^+ D^m X+V_X(X)+\frac{1}{4g_5^2}(F_L^2+F_R^2)),
\end{eqnarray}
with $\Phi$ the dilaton field, $X$ a complex scalar field, $V_X$ the scalar potential, $F_{mn}$ the field strength defined as $F^{L/R}_{mn}=\partial_m A^{L/R}_{n}-\partial_n A^{L/R}_{m}-i[A^{L/R}_{m},A^{L/R}_{,n}]$ in terms of the left/right hand gauge potential $A^{L/R}$, $g_5$ the gauge coupling, $g$ the determinant of metric $g_{mn}$, and the covariant derivative $D_m$  defined as $D_mX=\partial_mX-i A^L_mX+i XA^R_m$. The scalar potential $V_X$ only takes the mass term and has the form of
\begin{eqnarray}
V_X(X)=M_5^2 X^{+}X.
\end{eqnarray}
From the AdS/CFT prescription $M_5^2=(\Delta-p)(\Delta+p-4)$\cite{Witten:1998qj}, the mass of the complex scalar field $X$ $M_5^2$ can be determined as $M_5^2=\frac{-3}{L^2}$ (we will take the AdS radius $L=1$ in this work) by taking $\Delta=3, p=0$. The dilaton field is taken to be a simple quadratic form $\Phi(z)=\mu^2 z^2$. In this way, the meson spectra are shown to be linear with respect to the radial excitation quantum number $n$ at large $n$, which gives good description of the linear behavior of meson spectra. However, in the original soft-wall model,  there is no spontaneous chiral symmetry breaking in QCD vacuum and also no restoration at sufficient high temperature.

As pointed out in \cite{Chelabi:2015cwn,Chelabi:2015gpc}, further modifications of the dilaton field and the scalar potential are necessary in order to describe the spontaneous chiral symmetry breaking in QCD vacuum and its restoration.
The specific profile of the dilaton field are proposed \cite{Chelabi:2015cwn,Chelabi:2015gpc} and it takes the following form
\begin{eqnarray}\label{int-dilaton}
\Phi(z)=-\mu_1^2z^2+(\mu_1^2+\mu_0^2)z^2\tanh(\mu_2^2z^2),
\end{eqnarray}
which tends to be pure negative quadratic $\Phi(z)\simeq -\mu_1^2z^2+o(z^2)$ in ultraviolet(UV) region $z\rightarrow 0$ and positive quadratic form $\Phi(z)\simeq \mu_1^2 z^2$ in the infrared(IR) region $z\rightarrow \infty$. The positive quadratic behavior of $\Phi(z)$ is responsible for the linear spectra, which is well known in the soft-wall AdS/QCD.
The scalar potential takes the first several leading powers of $V_X$ and has the form of
\begin{eqnarray}
V_X(X)=M_5^2 X^{+}X+\lambda|X|^4+\gamma Re[det(X)].
\end{eqnarray}
In  \cite{Chelabi:2015cwn,Chelabi:2015gpc}, we have shown that the negative part as well as the quartic term $\lambda|X|^4$ in the scalar potential are essential for the spontaneous chiral symmetry breaking in the vacuum as well as its restoration at sufficient high temperature. In $SU(2)$ or two-flavor case, the t'Hooft determinant term $\gamma Re[det(X)]$ is taken to be zero, and we find a second order phase transition in the chiral limit and a crossover transition at any finite quark mass case. In $SU(3)$ or three-flavor case, we consider the t'Hooft limit, and we find a first order chiral phase transition in the chiral limit, while only at sufficient quark mass the phase transition turns to be a crossover one. Furthermore, in our recent study \cite{Li:2016gfn}, we show that after introducing magnetic field through the Einstein-Maxwell sector, the above $SU(2)$ model can describe inverse magnetic catalysis in the soft-wall model quite well. All the qualitative results are in agreement with the current understanding from lattice simulations\cite{qcd-phase-diagram,deForcrand:2006pv,Kanaya:2010qd,Bali:2011qj,Bali:2012zg} and other non-perturbative studies.

Nevertheless, in our previous study, we have only considered equal mass in both the $SU(2)$ and $SU(3)$ cases, i.e., $m_u=m_d$ and $m_u=m_d=m_s$. Therefore, we can only obtain the top line and the diagonal line in Fig.\ref{Colombia-Plot}. As shown in Fig.\ref{Colombia-Plot}, it is also interesting to consider the case $N_f=2+1$ with $m_l\equiv m_u=m_d\neq m_s$ case. It would be a quite natural extension of our previous study in $SU(3)$ case to the $N_f=2+1$ case.
In the following, $A_L, A_R$ will be set to be zero, since only the scalar field $X$ is relevant for the chiral phase transition.
Since $m_l\equiv m_u=m_d\neq m_s$, the complex scalar field should take the following form
\begin{eqnarray}
X=\left(
    \begin{array}{ccc}
      \frac{\chi_l(z)}{\sqrt{2}} & 0 & 0 \\
      0 & \frac{\chi_l(z)}{\sqrt{2}} & 0 \\
      0 & 0 &\frac{\chi_s(z)}{\sqrt{2}}\\
    \end{array}
  \right),
\end{eqnarray}
instead of a simple $\chi I_3$ with $I_3$ the $3\times3$ matrix. Here, we assume that $\chi_l(z),\chi_s(z)$ are functions of the 5D coordinate $z$ only and the factor $\frac{1}{\sqrt{2}}$ is just a normalization constant\footnote{We set this factor to make sure the coefficients of kinetic terms of $\chi_{u},\chi_{d},\chi_{s}$ are $\frac{1}{2}$. }. When $m_l\neq m_s$, we should have $\chi_l\neq \chi_s$, since the boundary values of $\chi_l,\chi_s$ are related to the quark masses. Under this ansatz, we can get the effective form of the action Eq.\ref{kkssaction} as following
\begin{eqnarray}\label{eff-action}
S[\chi_l,\chi_s]=-\int d^5x
 \sqrt{-g}e^{-\Phi}\left\{g^{zz}(\chi_l^{'2}+\frac{1}{2}\chi_s^{'2})+\frac{1}{L^2}\left[-3(\chi_l^2+\frac{1}{2}\chi_s^2)+v_4(2\chi_l^4+\chi_s^4)+3v_3\chi_l^2\chi_s\right]\right\},\nonumber\\
\end{eqnarray}
with $v_3=\frac{2\sqrt{2}}{3}\gamma, v_4=\frac{\lambda}{4}$ and $L$ the AdS radius, which will not affect the final results and will be taken to be $1$ later.

As in \cite{Chelabi:2015cwn,Chelabi:2015gpc}, we will neglect the back-reaction of $\chi_l, \chi_s$ to the background metric, and take the simple AdS-Schwarzchild(AdS-SW) black hole solutions
\begin{eqnarray}
dS^2&=&e^{2A_s(z)}(-f(z)dt^2+\frac{1}{f(z)}dz^2+dx_idx^i),\label{metric}\\
A_s(z)&=&-\log(z),\label{As}\\
f(z)&=&1-\frac{z^4}{z_h^4}\label{f},
\end{eqnarray}
where $z_h$ is the black hole horizon defined at $f(z_h)=0$ and could be related to the temperature $T$ by the following relation
\begin{eqnarray}
T=|\frac{f^{'}(z_h)}{4\pi}|=\frac{1}{\pi z_h}.\label{hawking-T}
\end{eqnarray}
Under this ansatz, the equations of motion for $\chi_l,\chi_s$ could be derived as the following form
\begin{eqnarray}
\chi_l^{''}+(3A_s^{'}-\Phi^{'}+\frac{f^{'}}{f})\chi_l^{'}+\frac{e^{2A_s}}{f}(3\chi_l-3v_3\chi_l\chi_s-4v_4\chi_l^3)&=&0,\label{eom-chil}\\
\chi_s^{''}+(3A_s^{'}-\Phi^{'}+\frac{f^{'}}{f})\chi_s^{'}+\frac{e^{2A_s}}{f}(3\chi_s-3v_3\chi_l^2-4v_4\chi_s^3)&=&0.\label{eom-chis}
\end{eqnarray}

Please notice that in the $SU(3)$ case with equal quark mass $m_l(\equiv m_u=m_d)=m_s$, we can have $\chi_l=\chi_s\equiv\chi$, and the above two equations will be reduced to the same one
\begin{eqnarray}
\chi^{''}+(3A_s^{'}-\Phi^{'}+\frac{f^{'}}{f})\chi^{'}+\frac{e^{2A_s}}{f}(3\chi-3v_3\chi^2-4v_4\chi^3)=0,
\end{eqnarray}
which is exactly the same as the one in \cite{Chelabi:2015cwn,Chelabi:2015gpc}. In \cite{Chelabi:2015cwn,Chelabi:2015gpc}, we have taken $v_3=-3, v_4=8$ to show the qualitative behavior of chiral phase transition in this model. Therefore, in this work we will continue to use this group of parameters.

The dilaton profile Eq.\ref{int-dilaton} has been shown to give well description of both chiral symmetry breaking and linear confinement. Thus in this work we will stick to this profile and extend the model to $m_l\neq m_s$ case. The parameters $\mu_0,\mu_1,\mu_2$ in the dilaton profile Eq.(\ref{int-dilaton}) will be taken as the same value
\begin{eqnarray}
\mu_0=0.43 \rm{GeV},\mu_1=0.83\rm{GeV}, \mu_2=0.176\rm{GeV}
\end{eqnarray}
as in \cite{Chelabi:2015cwn,Chelabi:2015gpc}. In the next section, we will show how to solve the order parameter of chiral phase transition from this model.

\subsection{Boundary condition and numerical solutions}
\label{sec-BDC}

Under the background Eqs.(\ref{int-dilaton},\ref{metric},\ref{As},\ref{f}), the equations of motion for $\chi_l, \chi_s$ could be solved numerically. Before that, we should specify the boundary condition.

Firstly, near the Ultraviolet (UV) boundary $z=0$, one can extract the perturbative expansion solution of $\chi_l, \chi_s$ as
\begin{eqnarray}
\chi_l&=&c_l z-3 c_l c_s v_3 z^2-(\mu_1^2-2c_l^2v_4+\frac{9}{2}c_s^2v_3^2+\frac{9}{2}c_l^2v_3^2)c_lz^3\log( z)+d_l z^3+...,\label{chiluv}\\
\chi_s&=&c_s z-3 c_l^2 v_3 z^2-(\mu_1^2-2c_s^2v_4-9 c_l^2v_3^2)c_s z^3\log(z)+d_s z^3+...,\label{chisuv}
\end{eqnarray}
with $c_l, d_l, c_s, d_s$ four integral constants of the two coupled second order ordinary derivative equations Eqs.(\ref{eom-chil},\ref{eom-chis}), which could be related to the current quark masses $m_l, m_s$ and chiral condensates $\sigma_l\equiv \langle\bar{u}u\rangle=\langle\bar{d}d\rangle, \sigma_s\equiv\langle\bar{s}s\rangle$ by the following equations\cite{Karch:2006pv,Cherman:2008eh}
\begin{eqnarray}
c_l&=&m_l \zeta,\\
d_l&=&\frac{\sigma_l}{ \zeta},\\
c_s&=&m_s \zeta,\\
d_s&=&\frac{\sigma_s}{\zeta},
\end{eqnarray}
and $\zeta=\frac{\sqrt{3}}{2\pi}$. As mentioned above, in this work we will try to study chiral phase transition under different values of $m_l, m_s$, so we will tune $m_l, m_s$ in the later calculation. At a first sight, the other two integral constants $\sigma_l,\sigma_s$ could be chosen independently on $m_l, m_s$. However, if one checks the two equations of motion Eqs.(\ref{eom-chil},\ref{eom-chis}), there are terms like \begin{eqnarray}
\frac{f^{'}\chi_l^{'}+e^{2A_s}(3\chi_l-v_3\chi_l\chi_s-v_4\chi_l^3)}{f(z)},\label{Ql}\\
\frac{f^{'}\chi_s^{'}+e^{2A_s}(3\chi_s-v_3\chi_l^2-v_4\chi_s^3)}{f(z)},\label{Qs}
\end{eqnarray}
with $f(z)$ in the denominator. At the horizon $z=z_h$, we have $f(z_h)=0$. Thus, $z_h$ is an apparent singularity of Eqs.(\ref{eom-chil},\ref{eom-chis}), where $\chi_l,\chi_s$ might be divergent. To avoid this divergence, physical solutions of $\chi_l,\chi_s$ should also satisfy the infrared (IR) boundary conditions
\begin{eqnarray}
Q_l(z_h)\equiv f^{'}\chi_l^{'}+e^{2A_s}(3\chi_l-v_3\chi_l\chi_s-v_4\chi_l^3)|_{z=z_h}=0,\label{chilir}\\
Q_s(z_h)\equiv f^{'}\chi_s^{'}+e^{2A_s}(3\chi_s-v_3\chi_l^2-v_4\chi_s^3)|_{z=z_h}=0,\label{chisir}
\end{eqnarray}
to  cancel the singularity at the horizon where $f(z_h)=0$.
It is easy to understand that with these two additional conditions from the requirement of the regularity of $\chi_l,\chi_s$, the other two UV coefficients $\sigma_l,\sigma_s$ cannot be considered as free integral constants with given $m_l, m_s$. Instead, they should be solved from the equations of motion.

\begin{figure}[h]
\begin{center}
\epsfxsize=6.5 cm \epsfysize=6.5 cm \epsfbox{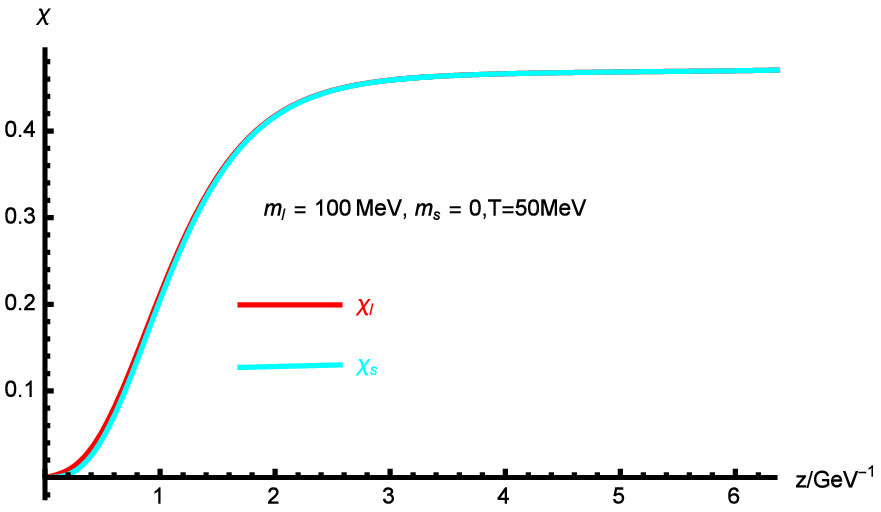}
\hspace*{0.1cm} \epsfxsize=6.5 cm \epsfysize=6.5 cm \epsfbox{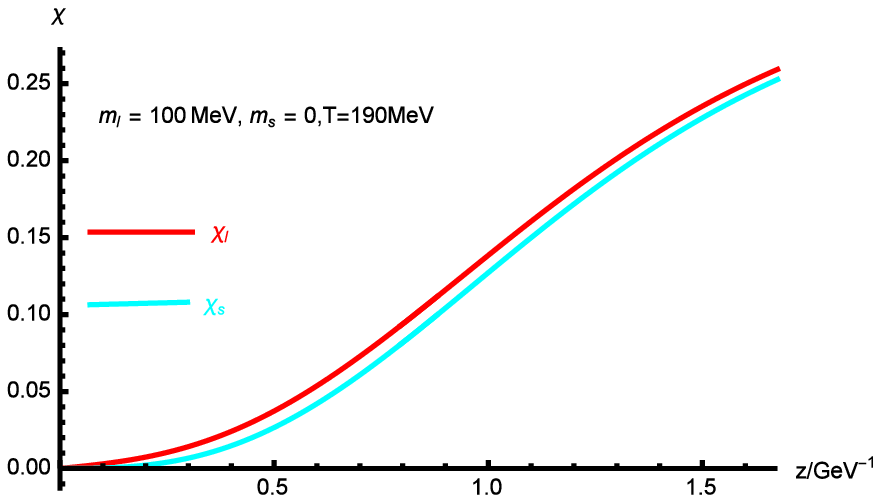}
\vskip -0.05cm \hskip 0.15 cm
\textbf{( a ) } \hskip 6.5 cm \textbf{( b )} \\
\end{center}
\caption{Solutions of $\chi_l$ and $\chi_s$ as functions of $z$ when $m_{l}=100{\rm MeV}, m_s=0$. Panel(a) shows the low temperature solutions with $T=50{\rm MeV}$ and Panel(b) shows the behavior for $T=190{\rm MeV}$. At low temperature, $\chi_l$ and $\chi_s$ are almost the same while at higher temperature they would be separated. }
\label{chi-z}
\end{figure}

Therefore, to find a physical solution with given $m_l,m_s$, one has to solve the boundary value problem
\begin{eqnarray}
& & \underset{\epsilon\to 0}{\lim}\frac{\chi_l(\epsilon)}{\epsilon}=m_l \zeta, ~~ \underset{\epsilon\to 0}{\lim}\frac{\chi_s(\epsilon)}{\epsilon}=m_s \zeta, \nonumber \\
& & Q_l(z_h)=0, ~~ Q_s(z_h)=0.
 \end{eqnarray}
 One can use the ``Shooting Method" to solve this boundary value problem. After it was solved, one can extract the chiral condensates $\sigma_l$ and $\sigma_s$. As an example, we take $m_{l}=100{\rm MeV}, m_s=0$ , $T=50{\rm MeV}, z_h=\frac{1}{\pi T}\approx 6.37 {\rm GeV}^{-1}$ and $T=190{\rm MeV}, z_h=\frac{1}{\pi T}\approx 1.67 {\rm GeV}^{-1}$. For $T=50{\rm MeV}, z_h=\frac{1}{\pi T}\simeq 6.37 {\rm GeV}^{-1}$, we get $\sigma_l=0.10 {\rm GeV}^3\approx(470{\rm MeV})^3, \sigma_s=0.12 {\rm GeV}^3\approx(495{\rm MeV})^3$ using ``Shooting Method" and plot the corresponding regular $\chi_l, \chi_s$ solutions in Fig.\ref{chi-z}(a). The non-vanishing values of $\sigma_l, \sigma_s$ at low temperature are signal of chiral symmetry breaking of the vacuum. From the figure, we could see that, at small $z$, $\chi_l, \chi_s$ are slightly separated from each other, since the leading terms in this region are $m_l \zeta z$ and $m_s \zeta z$, which are different when $m_l\neq m_s$. In the IR region when $z$ is large, $\chi_l, \chi_s$ is almost overlap and approach a constant value $\chi_l^h\equiv \chi_l(z_h)\approx \chi_s^h\equiv\chi_s(z_h)\approx 0.47$. Like in $m_l=m_s$ case, the UV region of the solutions are governed by the trivial vacuum $\chi_l=\chi_s=0$ while the IR region are governed by the non-trivial vacuum $\chi_l\neq0,\chi_s\neq0$.

Then at temperature up to $T=190{\rm MeV}$, we get $\sigma_l=0.06 {\rm GeV}^3\approx(389{\rm MeV})^3, \sigma_s=0.07 {\rm GeV}^3\approx(412{\rm MeV})^3$, which are smaller than the corresponding values at $T=50{\rm MeV}$, showing that chiral condensate are partly destroyed by temperature. In Fig.\ref{chi-z}(b) we plot the solutions of $\chi_l, \chi_s$ at temperature $T=190 {\rm MeV}$. From the figure, we could see that at high temperature, $\chi_l, \chi_s$ are still interpolation of the trivial vacuum and non-zero horizon values $\chi_l^h,\chi_s^h$. The separation of $\chi_l, \chi_s$ become larger than that at low temperature.

As a short summary, we have shown that the regular condition of $\chi_l,\chi_s$ would require the condensates $\sigma_l,\sigma_s$ as functions of quark masses $m_l,m_s$ and temperature $T$. Since chiral condensates are the order parameters of chiral phase transition, we will work out the quark mass and temperature dependence of $\sigma_l, \sigma_s$ to extract the information of chiral phase transition in next section.

\section{Chiral condensate and phase diagram in mass plane}
\label{massdiagram}

As mentioned in the introduction, it is also interesting to consider the property of chiral phase transition when $m_l\neq m_s$. In Sec.\ref{sec-BDC} we have shown that in the extended $N_f=2+1$ model one can solve chiral condensates $\sigma_l, \sigma_s$ from the equations of motion Eqs.(\ref{eom-chil},\ref{eom-chis}) with given quark masses $m_l, m_s$ and the temperature $T$. Hence, in this section we will try to extract the quark masses and temperature dependence of chiral condensates, which contains the information of chiral phase transition.

\begin{figure}[h]
\begin{center}
\epsfxsize=6.5 cm \epsfysize=6.5 cm \epsfbox{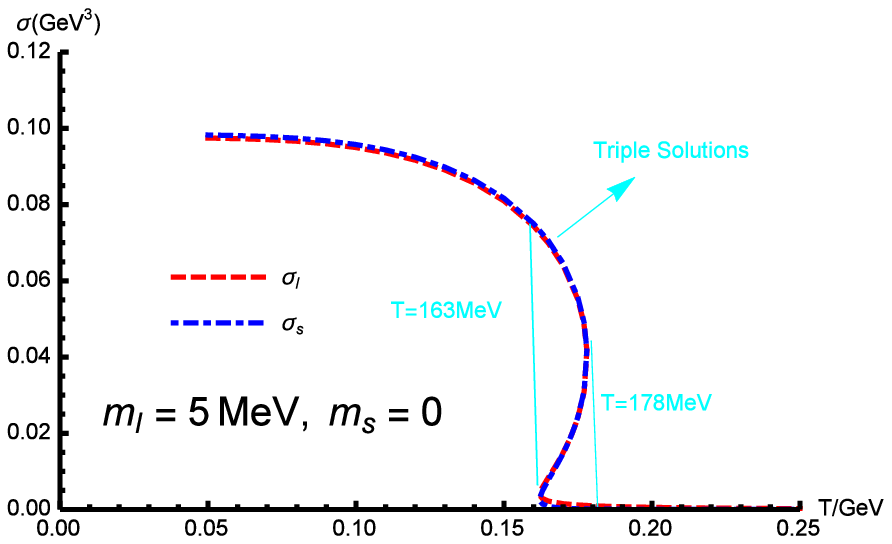}
\hspace*{0.1cm} \epsfxsize=6.5 cm \epsfysize=6.5 cm \epsfbox{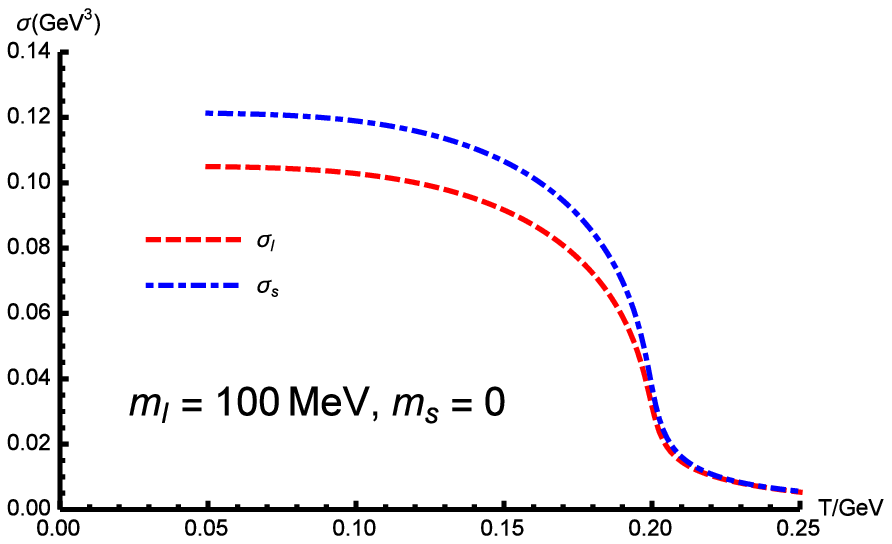}
\vskip -0.05cm \hskip 0.15 cm
\textbf{( a ) } \hskip 6.5 cm \textbf{( b )} \\
\end{center}
\caption{Chiral condensates $\sigma_l,\sigma_s $ as functions of temperature $T$. In Panel.(a), $\sigma_l, \sigma_s$ for $m_{l}=5{\rm MeV}, m_s=0$ are given. Below $T=163{\rm MeV}$ and above $T=178{\rm MeV}$, both $\sigma_l$ and $\sigma_s$ decrease monotonically with temperature $T$, while between $T=163{\rm MeV}$ and $T=178{\rm MeV}$ they are triple-value functions of $T$, showing a kind of characteristic behavior of first order phase transition.  In Panel.(b), the behavior of $\sigma_{l,s}$ for $m_{l}=100{\rm MeV}, m_s=0$ are given. A kind of monotonically decreasing in the whole temperature region are shown, which shows a characteristic behavior of crossover transition. }
\label{sigma-m}
\end{figure}

\begin{figure}[h]
\begin{center}
\epsfxsize=6.5 cm \epsfysize=6.5 cm \epsfbox{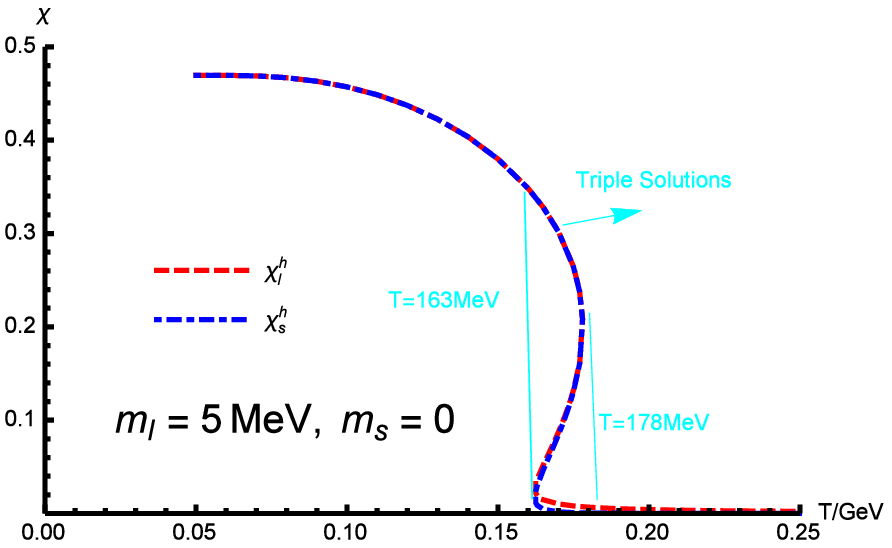}
\hspace*{0.1cm} \epsfxsize=6.5 cm \epsfysize=6.5 cm \epsfbox{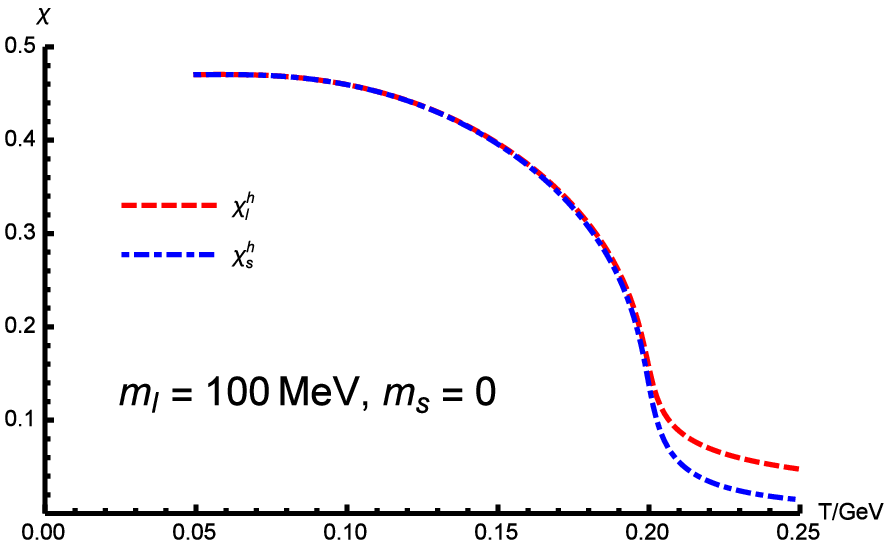}
\vskip -0.05cm \hskip 0.15 cm
\textbf{( a ) } \hskip 6.5 cm \textbf{( b )} \\
\end{center}
\caption{The near horizon boundary value $\chi_l^h, \chi_s^h$ as functions of temperature $T$. Panel.(a) and Panel.(b) give the result for $m_{l}=5{\rm MeV},m_s=0$ and $m_{l}=100{\rm MeV},m_s=0$ respectively. At low temperature, $\chi_l$ and $\chi_s$ are almost the same while above transition point they are separated.}
\label{chi-m}
\end{figure}

Firstly, we take $m_s=0$ and $m_l=5{\rm MeV}$. Using ``Shooting Method", we solve $\sigma_l,\sigma_s$ from Eqs.(\ref{eom-chil},\ref{eom-chis}). The results are shown in Fig.\ref{sigma-m}(a). From the figure, we could see that since $m_l=5{\rm MeV}\simeq m_s=0$, the differences of $\sigma_l$ and $\sigma_s$ are not very large. At low temperature, below $100{\rm MeV}$, both $\sigma_l$ and $\sigma_s$ are almost constants $0.1 {\rm GeV}^3$, showing the breaking of chiral symmetry in the vacuum. At high temperature, above $185{\rm MeV}$, $\sigma_l$ and $\sigma_s$ decrease to number smaller than $0.001{\rm GeV}^3$. As we showed in \cite{Chelabi:2015gpc}, the non-zero value of the high temperature tail comes from the non-zero quark masses other than spontaneous chiral symmetry breaking, since in the chiral limit, this tail will tend to be zero. Therefore, in fact the high temperature tail stands for the symmetry restoration phase. So from the numerical results in Fig.\ref{sigma-m}(a), we see a low temperature symmetry breaking phase and a high temperature symmetry restoration phase, which indicates a phase transition. Then, we look into the intermediate temperature region and found that within $163{\rm MeV}< T< 178{\rm MeV}$ there are three branches of solutions at the same temperature. As discussed in \cite{Chelabi:2015gpc}, this kind of behavior is a characteristic signal of first order phase transition. The exact transition temperature would located inside the temperature region $163{\rm MeV}< T< 178{\rm MeV}$ and can be worked out from the free energy. However, here we will focus on the order of the transition other than the critical temperature, so we would not try to extract the exact transition temperature for the first order transition. Furthermore, we also plot the temperature dependence of the horizon value $\chi_l^h\equiv \chi_s(z_h), \chi_s^h\equiv \chi_s(z_h)$ in Fig.\ref{chi-m}(a). There, we can see the same behavior as Fig.\ref{sigma-m}(a). At small temperature, $\chi_l,\chi_s$ are dominant by the non-trivial vacuum of scalar potential, while at high temperature they are dominant by the trivial $\chi_l=0,\chi_s=0$ vacuum.

Then, we increase $m_l$ to $m_l=100 {\rm MeV}$ while keeping $m_s=0$. After solving the equations of motion, the results of chiral condensates $\sigma_l, \sigma_s$ and $\chi_l^h,\chi_s^h$ are given in Fig.\ref{sigma-m}(b) and \ref{chi-m}(b), respectively. There we could see that at low temperature, both $\sigma_l$ and $\sigma_s$ increase with the increasing of $m_l$. Below $T=120{\rm MeV}$, $\sigma_l$ and $\sigma_s$ are almost constants $\sigma_l\approx 0.11{\rm GeV}^3, \sigma_s\approx 0.12{\rm GeV}^3$. It is also easy to see that $\sigma_s$ increase faster than $\sigma_l$. As a result, the separation of $\sigma_l$ and $\sigma_s$ becomes larger than that at $m_l=5{\rm MeV}, m_s=0$. At high temperature, again we could see that $\sigma_l,\sigma_s$ decrease to a very small value, showing the restoration of the spontaneous breaking symmetry(though explicit breaking is always there due to the non-zero quark masses). However, different from $m_l=5{\rm MeV}$ case, in this case $\sigma_l$ and $\sigma_s$ decrease monotonically from the vacuum expectation values to zero without the triple branches region. Furthermore, at around $T=200{\rm MeV}$, $\sigma_l$ and $\sigma_s$ decrease very fast from the value at symmetry breaking phase to the value at symmetry restoration phase. This kind of behavior shows a characteristic crossover phase transition.  As for $\chi_l^h,\chi_s^h$, we see that they also decrease monotonically from a larger value to zero. But differently, at low temperature, $\chi_l^h,\chi_s^h$ do not change much with the increasing of $m_l$. The low temperature value of these two quantities are still around $0.47$. At high temperature, they monotonically decrease to zero at different rate. From the figure, we could see that $\chi_l^h$ decreases faster than $\chi_s^h$.

\begin{figure}[h]
\begin{center}
\epsfxsize=6.5 cm \epsfysize=6.5 cm \epsfbox{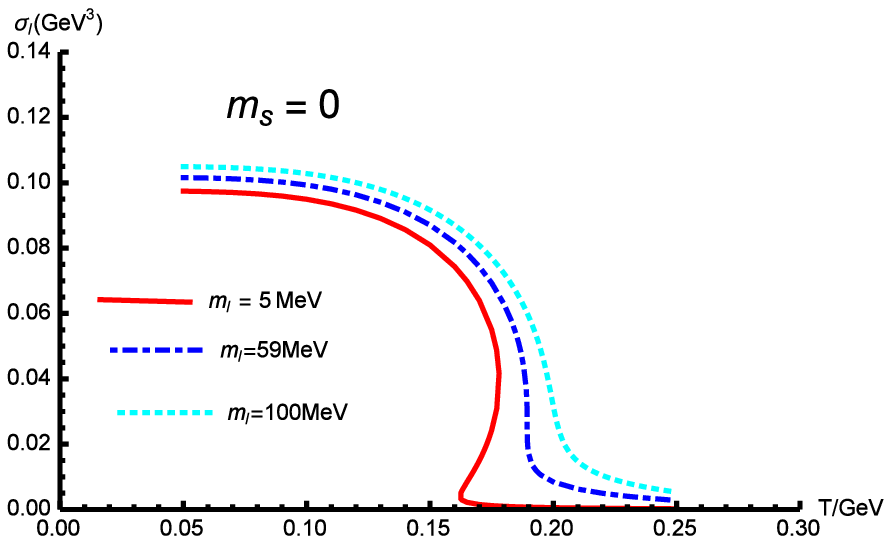}
\hspace*{0.1cm} \epsfxsize=6.5 cm \epsfysize=6.5 cm \epsfbox{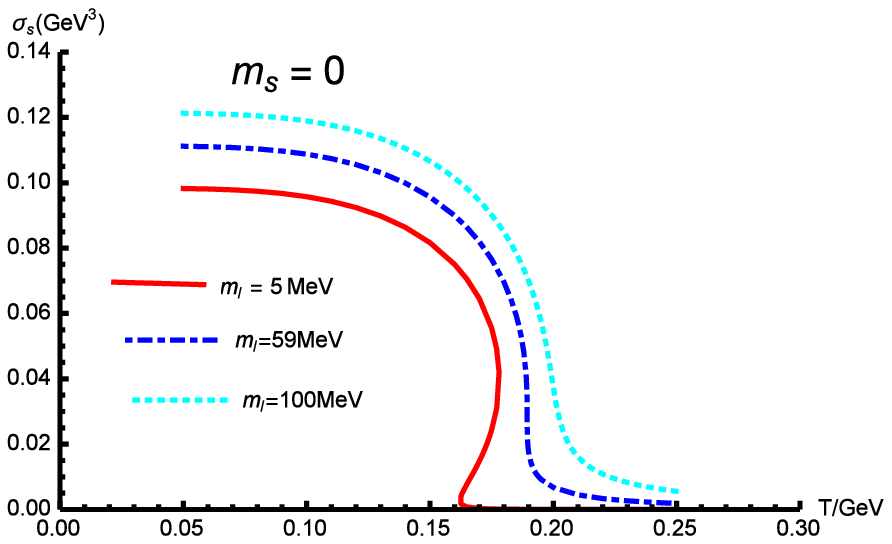}
\vskip -0.05cm \hskip 0.15 cm
\textbf{( a ) } \hskip 6.5 cm \textbf{( b )} \\
\end{center}
\caption{Panel.(a) shows the behavior of $\sigma_l$ as a function of temperature $T$ for different $m_{l}$ when $m_s=0$. Panel.(b) shows the behavior of $\sigma_s$ as a function of temperature $T$ for different $m_l$ when $m_s=0$. From Panel.(a,b), one can expect that when $m_s=0$ at small $m_{l}$ region the phase transition is of first order kind, while at sufficient high $T$ it turns to be a crossover one. The critical case happens when $m_{l}\approx 59{\rm MeV}$,  where $\frac{d\sigma}{dT}$ would diverge at around $T=189.3{\rm MeV}$ for both $\sigma_l$ and $\sigma_s$, showing a kind of second order phase transition. }
\label{sigma-ms=0}
\end{figure}

From the above discussion, it seems that when $m_s=0$, at small $m_l$ the system undergoes first order phase transition while at large $m_l$ the phase transition turns to be crossover. To be more rigorous, we fix $m_s=0$ and scan $m_l$. From Fig.\ref{sigma-ms=0}, we find that when $m_l$ is smaller than $59 {\rm MeV}$, both $\sigma_l,\sigma_s$ are non-monotonic, indicating a first order phase transition. Then the non-monotonic region shrinks as the increasing of $m_l$. At the critical value $m_l=59 {\rm MeV}$, the triple branches region disappears. At this value, we found that both the derivatives of $\sigma_l(T),\sigma_s(T)$ with respective to temperature $T$ diverge at the same temperature $T=189.3{\rm MeV}$, which reveals a second order phase transition. Then, above $m_l=59{\rm MeV}$, we find that $\sigma_l$ and $\sigma_s$ decreases monotonically from nonzero value to zero, showing a crossover phase transition.

\begin{figure}[h]
\begin{center}
\epsfxsize=6.5 cm \epsfysize=6.5 cm \epsfbox{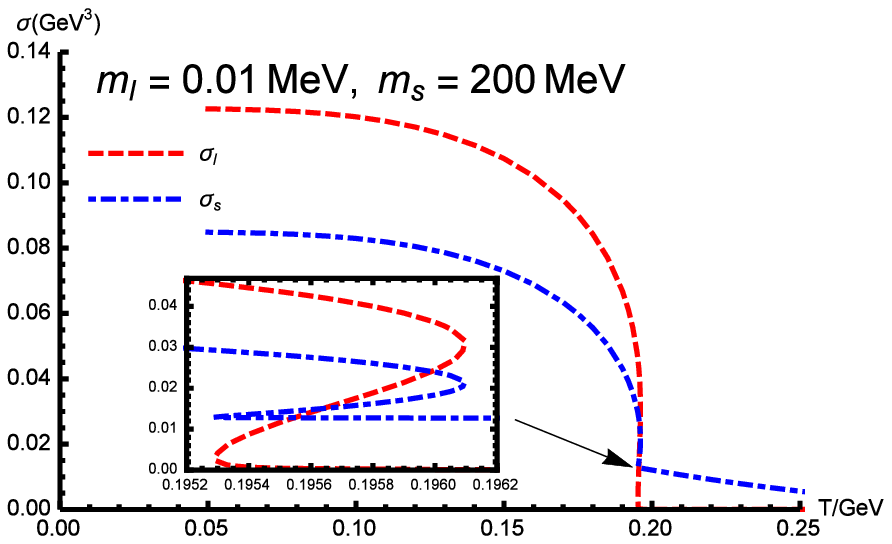}
\hspace*{0.1cm} \epsfxsize=6.5 cm \epsfysize=6.5 cm \epsfbox{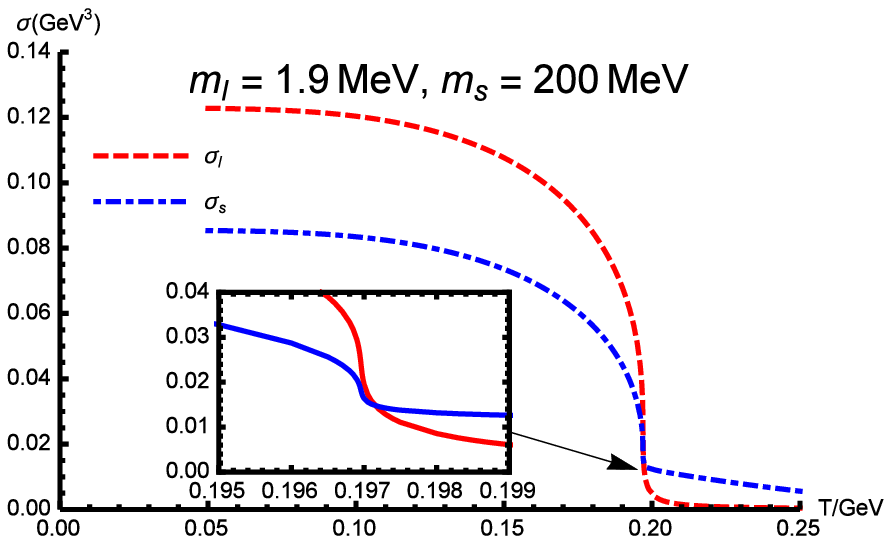}
\vskip -0.05cm \hskip 0.15 cm
\textbf{( a ) } \hskip 6.5 cm \textbf{( b )} \\
\end{center}
\caption{Panel.(a) shows the behavior of $\sigma_l$ as a function of temperature $T$ for different $m_{l}$ when $m_s=200{\rm MeV}$. Panel.(b) shows the behavior of $\sigma_s$ as a function of temperature $T$ for different $m_{l}$ when $m_s=200{\rm MeV}$. From Panel.(a,b), one can expect that when $m_s=200{\rm MeV}$ at small $m_{l}$ region the phase transition is of first order kind, while at sufficient high $T$ it turns to be a crossover one. The critical case happens when $m_{l}\approx 1.9{\rm MeV}$,  where $\frac{d\sigma}{dT}$ would diverge at around $T=196.9{\rm MeV}$ for both $\sigma_l$ and $\sigma_s$, showing a kind of second order phase transition.  }
\label{sigma-ms=200}
\end{figure}

Then we increase $m_s$ to $m_s=200{\rm MeV}$ and scan $m_l$. We plot the results of $\sigma_l, \sigma_s$ in Fig.\ref{sigma-ms=200}. We see that qualitatively the results are similar to those when $m_s=0$. When temperature increases, condensates would decrease from nonzero value to zero. When $m_l=0.01 {\rm MeV}$, there is a short region where $\sigma_l,\sigma_s$ have triple branches. But the triple solutions region becomes very short(from $T\approx 195.3{\rm MeV}$ to $T\approx 196.1{\rm MeV}$) comparing to $m_s=0$ case. Then at larger $m_l$ the behavior becomes crossover. The critical value of $m_l$ is around $1.9 {\rm MeV}$ and from Fig.\ref{sigma-ms=200}(b) the second order transition temperature becomes $196.9 {\rm MeV}$.

\begin{figure}[h]
\begin{center}
\epsfxsize=6.5 cm \epsfysize=6.5 cm \epsfbox{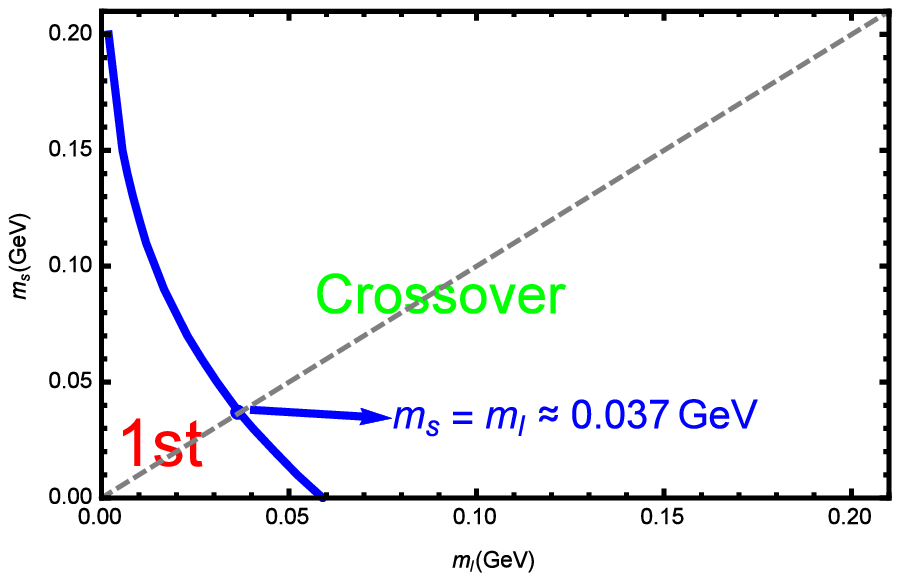}
\hspace*{0.1cm} \epsfxsize=6.5 cm \epsfysize=6.5 cm \epsfbox{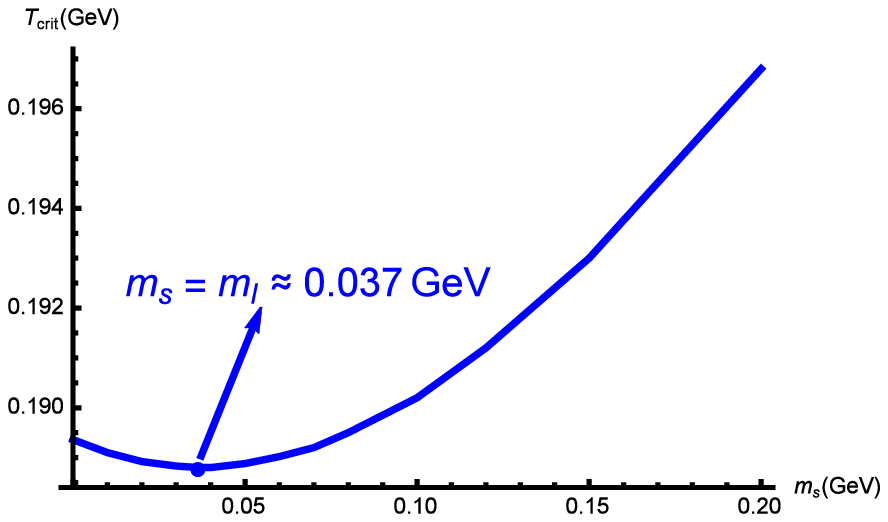}
\vskip -0.05cm \hskip 0.15 cm
\textbf{( a ) } \hskip 6.5 cm \textbf{( b )} \\
\end{center}
\caption{The phase diagram for chiral phase transition in $m_l-m_s$ plane. The blue solid line in Panel.(a) shows the critical line(second order line) between the first order region(the bottom left corner) and the crossover region(the upper right corner). In Panel.(b), the temperature for the critical line in Panel.(a) are given. The blue dot in Panel.(a)(b) are $m_s=m_l=0.037{\rm GeV}$, the critical masses when $m_l=m_s$, which is the same as the one we extracted in $SU(3)$ case in \cite{Chelabi:2015gpc}. In Panel.(b), in the left branch to the blue dot, the transition temperature of the critical line decreases with the increasing of $m_s$, while in the right branch it increases, indicating that the two branches might be governed by different universality classes. }
\label{diagram-Nf2+1}
\end{figure}

Furthermore, we tune $m_s$ from $0$ to $200 {\rm MeV}$. We find that for each value of $m_s$, there is a critical $m_l$, at which chiral phase transition becomes second order. Here we note that both at large $m_s$ and small $m_s$, the triple solutions region of $\sigma_l,\sigma_s$ disappears at the same critical mass $m_l^c$ and $\frac{d\sigma}{dT}$ diverges at the same critical temperature for both $\sigma_l$ and $\sigma_s$. Below this critical $m_l^c$, the transition is of first order while above it the phase transition is crossover. We plot the critical line in Fig.\ref{diagram-Nf2+1}(a). From the figure, we see that the critical line divides the whole plane into two parts: the bottom left part is first order phase transition region while the upper right part is crossover transition region. Qualitatively, this result is in agreement with the ``Colombia Plot" in Fig.\ref{Colombia-Plot}(a), which is summarized from lattice simulations and other effective methods. Moreover, from the critical line, one can extract the exact transition temperature, where $\frac{d\sigma}{dT}$ diverges for both $\sigma_l$ and $\sigma_s$. The results are given in Panel.(b). There one can read that the transition temperature of the second order phase transition decreases when it approaches $m_l=m_s\approx0.037{\rm GeV}$ from left, while it increases when from right. This might indicate that the two branches of the critical line separated by the $m_l=m_s$ point might be governed by different universality classes, though the exact correspondence is out of the scope of this work. Roughly speaking, this is also consistent with Fig.\ref{Colombia-Plot}(a), where the upper part of the critical line is governed by $O(4)$ classes and the bottom part is governed by $Z(2)$ classes.

\section{Conclusion and discussion}
\label{sum}

To study QCD phase transitions in different situations are of great importance and it is interesting to consider chiral phase transition at different quark masses. In \cite{Chelabi:2015cwn,Chelabi:2015gpc}, we proposed a modified soft-wall AdS/QCD model and study chiral phase transition in $N_f=2$ and $N_f=3$ case. After extracting temperature dependent chiral condensate, it is found that chiral phase transition is of second order in two flavor chiral limit and turns to be crossover at any finite quark masses, while in three flavor chiral limit it becomes first order and turns to crossover only at sufficient large quark masses. Then, in \cite{Li:2016gfn}, we have shown that this model can describe inverse magnetic catalysis after considering the Einstein-Maxwell sector. Therefore, in this work, we try to extend these studies to $N_f=2+1$ case when $m_u=m_d\neq m_s$.

The extension of previous study is quite natural and simple. The main different is that the expectation value of the scalar field $X$ in soft-wall model should be taken as $3\times3$ diagonal matrix $diag\{\chi_l,\chi_l,\chi_s\}$ other than $2\times2$ $diag\{\chi,\chi\}$. If $m_{l}\neq m_s$, then it is expected that $\chi_l\neq \chi_s$ and one has to deal with the two coupled second order derivative equations. The UV boundary condition of $\chi_l,\chi_s$ can be related to quark masses $m_l,m_s$ and chiral condensates $\sigma_l, \sigma_s$. The black hole horizon would come up with another boundary condition, which will require condensates as functions of temperature and quark masses, i.e. of the form $\sigma_l(m_l,m_s,T),\sigma_s(m_l,m_s, T)$.

Fixing $m_s$ and solving the equations of motion, it is found that at both small $m_l$ and large $m_l$, chiral condensate would decrease from finite value at low temperature to zero at high temperature, indicating a phase transition between symmetry breaking phase at low temperature and symmetry restoration phase at high temperature. Moreover, it is found that at small $m_l$, $\sigma_l, \sigma_s$ are triple valued in certain temperature range, giving the signal of first order phase transition. The triple valued temperature range would decrease with the increasing of $m_l$, and at certain critical value it disappears and the phase transition become a second order one. If one continues to increase $m_l$, then $\sigma_{l},\sigma_s$ will decrease monotonically and the phase transition becomes crossover. Varying $m_s$, the qualitative behavior is similar. Thus, the whole $m_l-m_s$ plane is divided into two regions: first order region and crossover region as shown in Fig.\ref{diagram-Nf2+1}(a). The boundary of these two regions is the second order line, which could be extracted by solving critical values of $m_l$ at different $m_s$. The second order line are divided by the $m_l=m_s$ line into two parts, and it is shown that the $m_s$ dependence of the transition temperature in these two parts are totally contrast, which might indicate that the two parts are governed by different universality classes. Qualitatively, these model results for chiral phase transition  in Fig.\ref{diagram-Nf2+1} is in agreement with the ``Colombia Plot" in Fig.\ref{Colombia-Plot}(a) summarized from lattice simulations and other non-perturbative analysis. It confirms that the soft-wall AdS/QCD framework can provide good holographic description on chiral dynamics.

\vskip 0.5cm
{\bf Acknowledgement}
\vskip 0.2cm
The authors thank Kaddour Chelabi, Zheng Fang, Song He and Yue-Liang Wu for valuable discussions. M.H. is supported by the NSFC under Grant Nos. 11175251 and 11275213, DFG and NSFC (CRC 110),
CAS key project KJCX2-EW-N01, and Youth Innovation Promotion Association of CAS. This work is partly supported by China Postdoctoral Science Foundation(2015M580136).

\end{document}